\documentclass{svproc}
\usepackage{url}

\usepackage{amsmath,amssymb}
\usepackage{caption,subcaption}
\usepackage{graphicx,xcolor} % Required for inserting images
\usepackage{braket}
\newcommand{\La}{\mathcal{L}}

\begin{document}
\mainmatter              % start of a contribution
\title{Neutrino Oscillation in Core Collapse Supernova: The Impact of Spacetime Geometry}
\titlerunning{Neutrino in CCSN : Spacetime Geometry interaction}  % abbreviated title (for running head)
%\author{Indrajit Ghose\inst{1} \and Amitabha Lahiri\inst{2}}
\author{Indrajit Ghose \and Amitabha Lahiri}
\authorrunning{Indrajit Ghose and Amitabha Lahiri} % abbreviated author list (for running head)
\tocauthor{I.~Ghose and A.~Lahiri}
\institute{S. N. Bose National Centre for Basic Sciences
Block JD, Sector 3, Salt Lake,\\ WB 700106, INDIA,\\
\email{ghose.meghnad@gmail.com, amitabha@bose.res.in}\\
{\em\small Presented by Indrajit Ghose at the XXVI DAE-BRNS HEP Symposium,\\ 19-23 Dec 2024, Varanasi}
}
% \and
% S. N. Bose National Centre for Basic Sciences
% Block JD, Sector 3, Salt Lake, WB 700106, INDIA,\\
% \email{amitabha@bose.res.in}}

\maketitle              % typeset the title of the contribution

\begin{abstract}
Neutrino flavor evolution inside a core-collapse supernova is a topic of active research. The core of a supernova is an intense source of neutrinos and antineutrinos. Self-interaction among neutrinos (as well as antineutrinos) gives rise to a rich phenomenology not seen in terrestrial situations. In studies of the dynamics of flavor evolution in such environments, the gravitational effects are generally ignored. Although the curvature outside a dense core does not deviate much from a flat space, the spin of the neutrinos can still couple to the torsion of the spacetime. These extra degrees of freedom of curved spacetime have interaction strengths that are proportional to the density of the neutrinos and the other fermions \cite{Chakrabarty:2019cau} \cite{Barick:2023qjq} as well as the coupling constants of the spin-torsion interaction. We have studied the effects of such interactions in flavor evolution inside a core-collapse supernova \cite{Ghose:Manuscript}. The self-interaction gets modified by the spin-torsion interaction and the oscillation dynamics is modified. We have seen that there are noticeable changes in the flavor dynamics when the neutrino density is uniform. We have also studied the effects of such interaction in a realistic core-collapse supernova (CCSN). As neutrino astronomy enters the precision era, this study will shed light on the potential of neutrino fluxes from CCSN to probe the neutrino-neutrino interaction.
\end{abstract}
\section{ Spacetime Geometry couples with spin} In Einstein's description, gravity is described by the propagation of particles along a geodesic in curved spacetime. The components of the  affine connection defining the geodesics are symmetric. However, Cartan later expanded the theory by relaxing the symmetry constraint in the lower two indices of the connection components~\cite{Cartan:1923zea}. To form our theory of propagation of fermions in curved spacetime, we introduce the spin connection $A_{\mu}{}^{ab}=\omega_{\mu}{}^{ab}+S_{\mu}{}^{ab}$\,, where $\omega_{\mu}{}^{ab}$ corresponds to the torsion-free Levi-Civita connection which appears in Einstein's General Relativity (GR) and $S_{\mu}{}^{ab}$ is the contortion that appears due to the relaxation of symmetry of the connection. Greek indices will refer to the curved spacetime and the Latin indices will refer to the locally flat coordinate system. The minimal substitution to make the Dirac Lagrangian density as a scalar density under general coordinate transformation is
\begin{equation}
    \partial_{\mu}\psi \to D_{\mu}\psi = \partial_{\mu}\psi - (i/4)A_{\mu}{}^{ab}\sigma_{ab}\psi \label{eq:minimal_substituion}
.\end{equation}
Adding the Dirac Lagrangian after applying the aforementioned minimal substitution with the Geometrical Lagrangian density, the most generic solution of the contortion is~\cite{Chakrabarty:2019cau}
\begin{align}\label{chiral.torsion}
	S_{\mu}{}^{ab} &= (\kappa/4)\epsilon^{abcd}e_{c\mu} \sum\limits_i \left(-\lambda^i_{L}\bar{\psi}_{iL}\gamma_d \psi_{iL} + \lambda^i_{R}\bar{\psi}_{iR}\gamma_d \psi_{iR}\right)\,.
\end{align}
The $S_{\mu}{}^{ab}$ are non-dynamical and can be substituted back into the action. Assuming that the interaction violates parity maximally we get  
\begin{align}
	\La_\psi = &~\text{Dirac eqaution in GR} - \frac{1}{\sqrt{2}}\biggl(\sum_{i} \lambda_i \bar{\psi}_{iL}\gamma_d \psi_{iL}\biggr)^2\, \label{eq:spin-torsion_interaction}
.\end{align}
We have an effective torsionless theory with a quartic interaction term. The coupling constants $\lambda_{i}$ have a mass-dimension $-1$.\\
\vspace{-2em}

\section{Flavor evolution in the presence of self-interaction} In an effective two-neutrino family paradigm, the neutrinos can be expressed by a $2 \times 2$ Hermitian density matrix. The Hermitian matrices can be expressed in terms of the Pauli matrices
%write in one line
\begin{align}
\rho&=n/2(\mathbb{I}_2+\vec{P}\cdot \vec{\sigma}); &
\bar{\rho}&=\bar{n}/2(\mathbb{I}_2+\vec{\bar{P}}\cdot \vec{\sigma})\,. \label{eq:density_matrix}
\end{align}
$n, \bar{n}$ are the total density of neutrinos and antineutrinos. The equations of motion for $\vec{P}$ and $\vec{\bar{P}}$\,,
\begin{align}
\partial_{\tau}\vec{P}&=\biggl(\left(\hat{\omega}+\sqrt{2} agr R_f+\sqrt{2}R_{\nu}gr^2\vec{B}\cdot(\vec{P}-\vec{\bar{P}})\right)\vec{B}+\sqrt{2}R_e\vec{L}-\sqrt{2}R_{\nu}f_{g,r}\vec{\bar{P}}\biggr)\times\vec{P} \label{eq:P_eqn_red} \\
\partial_{\tau}\vec{\bar{P}}&=\biggl(\left(-\hat{\omega}+\sqrt{2}agr R_f+\sqrt{2}R_{\nu}gr^2\vec{B}\cdot(\vec{P}-\vec{\bar{P}})\right)\vec{B}+\sqrt{2}R_e\vec{L}+\sqrt{2}R_{\nu}f_{g,r}\vec{P}\biggr)\times\vec{\bar{P}} \label{eq:Pbar_eqn_red}
.\end{align}
Here we have defined $\hat{\omega}$, which parametrizes the mass hierarchy: $\hat{\omega}$ takes the value of $\pm 1$ for normal (inverted) hierarchy. Let us denote $\lambda_1^2=gG_{F}$\,, $\lambda_2=(2r+1)\lambda_1$\,, and $\lambda_f=a\lambda_1$\,, for some $g, r,$ and $a$ (we assume that $\lambda_1$ is non-vanishing). Assuming that the total density of neutrinos is the same as that of antineutrinos. Here we have introduced the dimensionless time $\tau=\frac{t \Delta m^2}{2E}$\,, the dimensionless reduced density parameters $R_e=\frac{2G_Fn_eE}{\Delta m^2}$\,, $R_f=\frac{2G_Fn_fE}{\Delta m^2}$\,, and $R_{\nu}=\frac{2G_FnE}{\Delta m^2}$\,, and also written $f_{g,r}=1+\frac{g}{4}(2r+1)$. Here $\lambda_{1,2}$ are the spin-torsion coupling constants for mass eigenstates $\ket{\nu_{1,2}}$\,, while $n_{e,f}$ are the electron and total non-neutrino fermion number density.
%%%%%%%%%%%%%%%%%%%%%%%%%%%%%%
\section{ Oscillation pattern} We have numerically solved the Eqs.~\eqref{eq:P_eqn_red} and \eqref{eq:Pbar_eqn_red}. For illustration, we use the neutrino oscillation parameters and relevant neutrino and matter densities as given in ~\cite{Lin:2022dek}. We assume a background populated by neutrinos with a single energy $E=15.1$ MeV, $\Delta m^2 = 2.5\times 10^{-3}$ eV$^2$ and the mixing angle is $8.6^\circ$. To express the densities of neutrinos and other relevant densities we introduce a number $\mu_0=1.76\times 10^5$, in units of which the reduced density parameters will be expressed. In a neutral background of electron fraction $0.5$ -- protons, neutrons, and electrons will have the same density. Hence, there will be 3 up and 3 down quarks for every electron. So we can write $R_f=7R_e$. In all these plots, the time axis will be marked as $\tau$\, which is time in units of $2E/\Delta m^2 = 8.3~\mu$s.
\vspace{-1em}
%subsections delete, only bold letters and colons
\begin{figure}[!ht]
     \centering
     \begin{subfigure}[t]{0.46\textwidth}
         \centering
         \includegraphics[width=\textwidth]{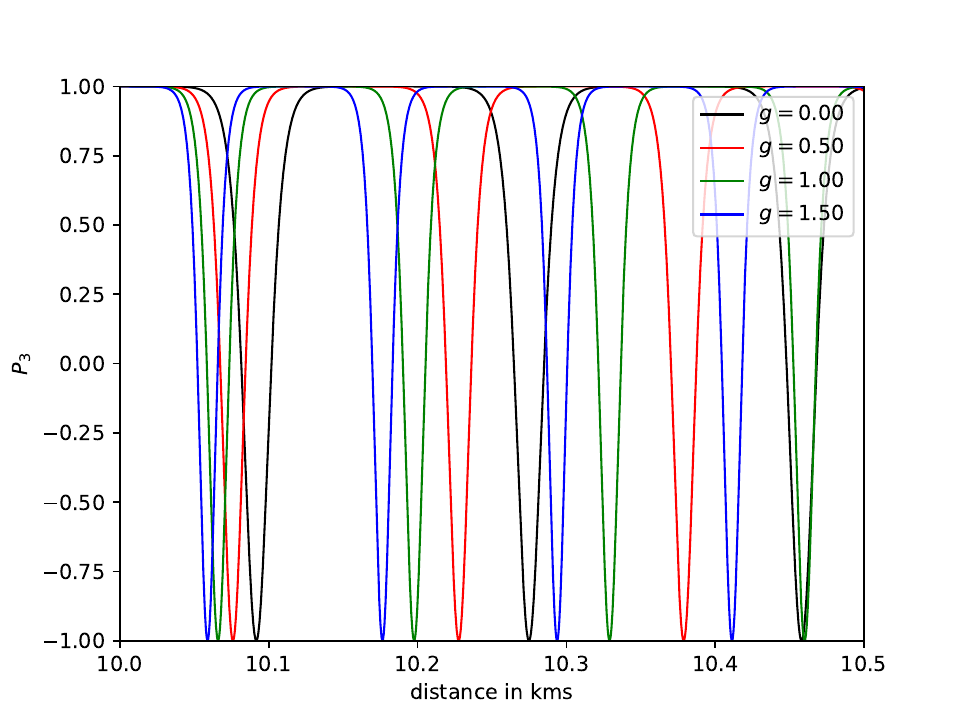}
         \caption{Evolution of $P_3$ when $a=0$.}
         \label{fig:simple_g_vary_a_0.0}
     \end{subfigure}
     \hfill
     \begin{subfigure}[t]{0.46\textwidth}
         \centering
         \includegraphics[width=\textwidth]{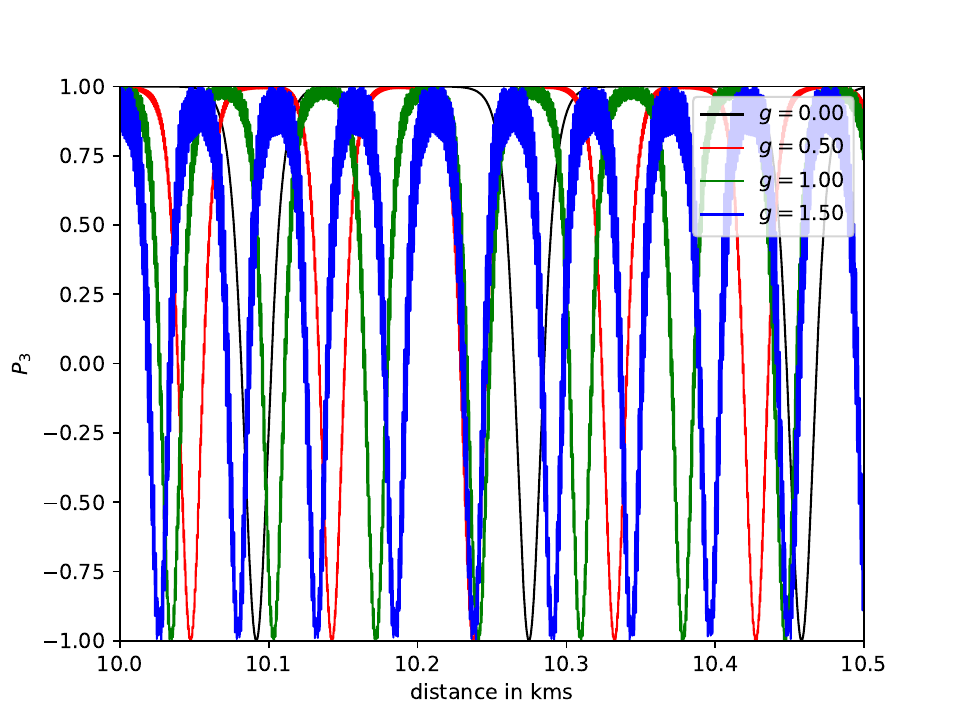}
         \caption{Evolution of $P_3$ when $a=0.1$.}
         \label{fig:simple_g_vary_a_0.1}
     \end{subfigure}
        \caption{\small{Oscillation pattern for uniform neutrino density. Both of these scenarios include $(R_{\nu},R_{e})=(\mu_0/10,\mu_0/10)$.}
        \label{fig:simple_g_vary}}
\end{figure}
\vspace{-3em}
\begin{figure}[!ht]
     \centering
     \begin{subfigure}[b]{0.46\textwidth}
         \centering
         \includegraphics[width=\textwidth]{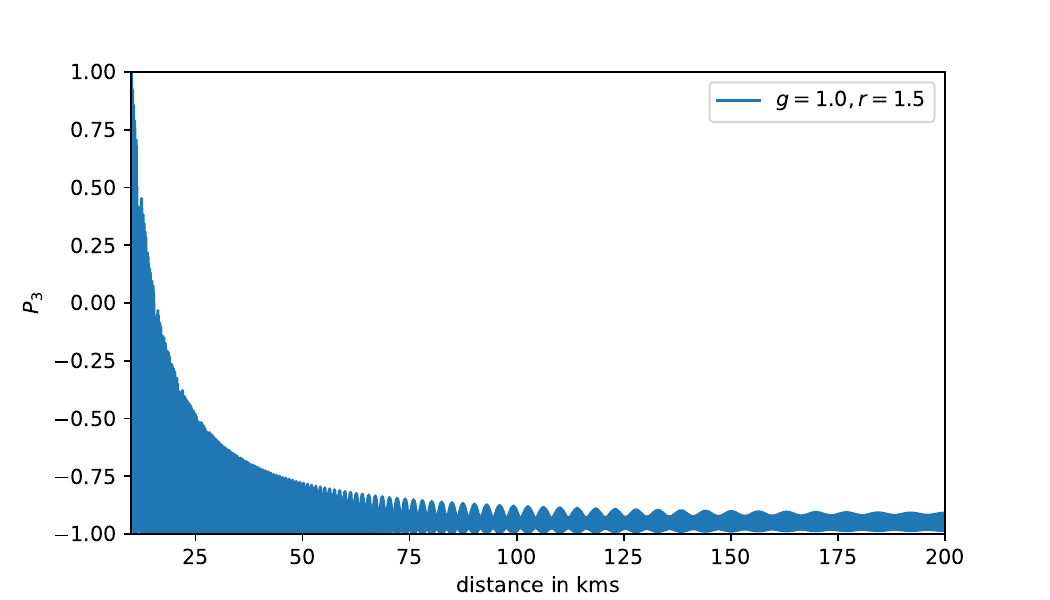}
         \caption{Evolution of $P_3$ for $r = 1.5$.}
         \label{fig:const-g_1.0_r_1.5_a_0.1}
     \end{subfigure}
     \hfill
     \begin{subfigure}[b]{0.46\textwidth}
         \centering
         \includegraphics[width=\textwidth]{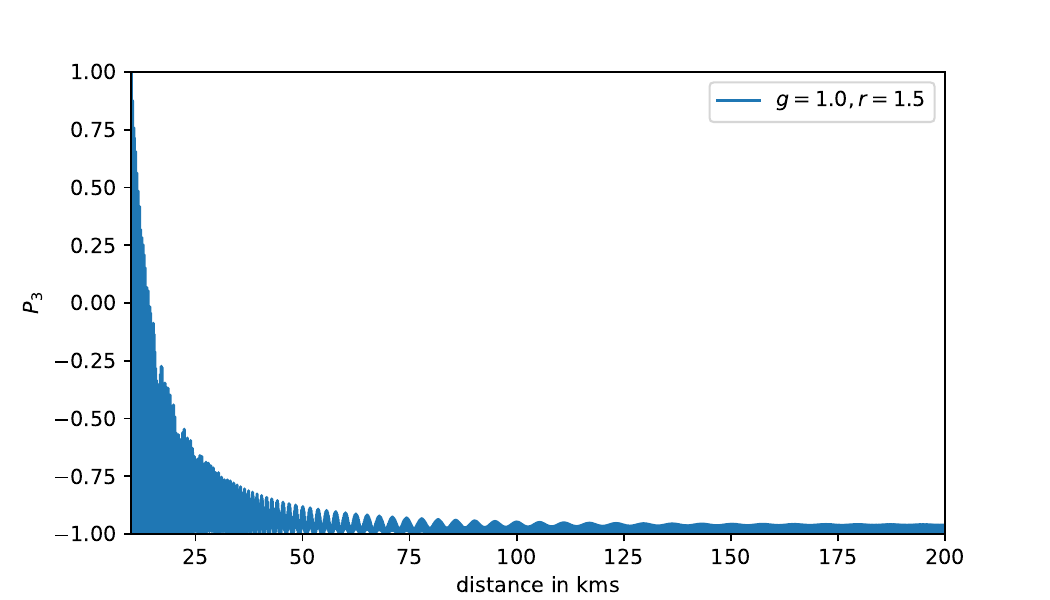}
         \caption{Evolution of $P_3$ for $r = -1.5$.}
         \label{fig:sconst-g_1.0_r_-1.5_a_0.1}
     \end{subfigure}
        \caption{\small{Oscillation pattern for non-uniform neutrino density. Both of these scenarios include $(R_{\nu} (R),R_{e})=(\mu_0/10,\mu_0/10)$. We have chosen $g=1.0,~a=0.1$ for the two plots. $R_e$ is uniform.}
        \label{fig:non-uniform_g_vary}}
\end{figure}
%\vspace{-2em}
%%%%%%%%%%%%%%
\\
\noindent{\bf Uniform density of neutrinos:} Flavor oscillation in the presence of uniform neutrino and fermion density in the background does not cause any permanent flavor change. The complete flavor oscillation gets fully transformed back into the initial $e$ neutrino. This behavior is seen clearly in Fig. \ref{fig:simple_g_vary}. Higher $g$ causes the first dip in the $P_3$ to move towards the left in Fig. \ref{fig:simple_g_vary_a_0.0}. Increased spin-torsion interaction induces flavor instability. The features are also present in Fig. \ref{fig:simple_g_vary_a_0.1}.\\
%\vspace{-2em}
%\vspace{-1em}

\begin{figure}[!hb]
	\centering
	\includegraphics[width=0.5\linewidth]{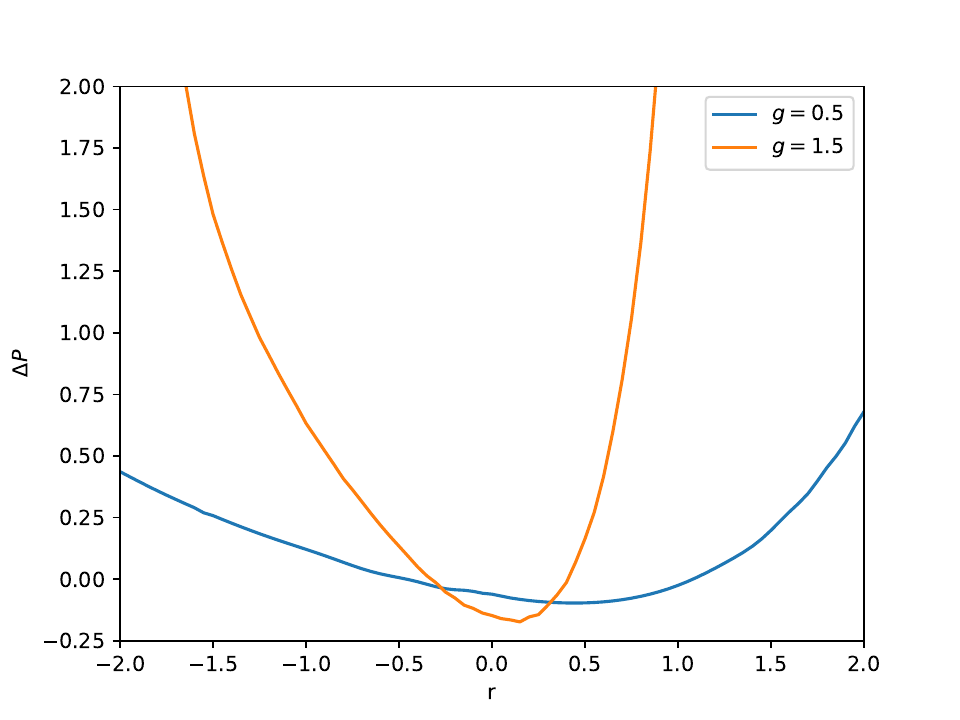}
	\caption{\small{Fractional change in survival probability away from the core as a function of $r$.}}
	\label{fig:del-P_r_vary}
\end{figure}

\noindent{\bf  Non-uniform density:} Outside the core of a realistic supernova, the neutrino number density falls with distance from the core ($d$) like~\cite{Duan:2006an}
\begin{align}
    R_{\nu,\bar{\nu}}(d)=R_{\nu,\bar{\nu}}(R)\left(1-\sqrt{1-R^2/d^2}\right)R^2/d^2
    \label{eq:neutrino_profile}
.\end{align}
$R$ is the radius of the proto-neutron star. We will use the same values of $\theta,~\Delta m^2,~E$ as the previous section. We will assume $R=10$ km. In the presence of neutrino density falling towards outside neutrino flavors change permanently in Fig. \ref{fig:non-uniform_g_vary}. Before ending, we display Fig.~\ref{fig:del-P_r_vary}, which shows the fractional change in survival probability $\Delta P(g,r)$,
\begin{align}
    \Delta P(g,r) %=(\mathcal{P}_{S}(g,r)-\mathcal{P}_{S}(0,0))/P_{S}(0,0)
    =(P_{\infty}(g,r)-P_{\infty}(0,0))/(1+P_{\infty}(0,0)) 
    \label{def:del_P}
.\end{align}
Here, $P_{\infty}$ defines the value of $P_3$ far away from the core.


\end{document}